\documentclass[journal,twoside]{IEEEtran}

\usepackage{cite}
\usepackage{amsmath,amssymb,amsfonts}
\usepackage{algorithmic}
\usepackage{graphicx}
\usepackage{textcomp}
\usepackage{xcolor}
\usepackage[caption=false]{subfig}
\usepackage{optidef}
\usepackage{booktabs}       
\newtheorem{property}{Property}
\newtheorem{definition}{Definition} 

\title{Neural optimization for quantum architectures: graph embedding problems with Distance Encoder Networks
}

\author{\IEEEauthorblockN{Chiara Vercellino\IEEEauthorrefmark{1}\IEEEauthorrefmark{2}\IEEEauthorrefmark{3},
Giacomo Vitali\IEEEauthorrefmark{1}\IEEEauthorrefmark{2},
Paolo Viviani\IEEEauthorrefmark{1},
Alberto Scionti\IEEEauthorrefmark{1},
Andrea Scarabosio\IEEEauthorrefmark{1},
Olivier Terzo\IEEEauthorrefmark{1},\\
Edoardo Giusto\IEEEauthorrefmark{2},
Bartolomeo Montrucchio\IEEEauthorrefmark{2}}\\
\IEEEauthorrefmark{1}\textit{LINKS Foundation}, Torino, Italy \\
\IEEEauthorrefmark{2}\textit{\textit{DAUIN}, Politecnico di Torino}, Torino, Italy\\
\IEEEauthorrefmark{3}\textit{chiara.vercellino@linksfoundation.com}
}

\begin{document}
\maketitle

\begin{abstract}
Quantum machines are among the most promising technologies expected to provide significant improvements in the following years. However, bridging the gap between real-world applications and their implementation on quantum hardware is still a complicated task. One of the main challenges is to represent through \textit{qubits} (\textit{i.e.}, the basic units of quantum information) the problems of interest. According to the specific technology underlying the quantum machine, it is necessary to implement a proper representation strategy, generally referred to as \textit{embedding}. This paper introduces a neural-enhanced optimization framework to solve the constrained unit disk problem, which arises in the context of qubits positioning for neutral atoms-based quantum hardware. The proposed approach involves a modified autoencoder model, \textit{i.e.}, the Distances Encoder Network, and a custom loss, \textit{i.e.}, the Embedding Loss Function, respectively, to compute Euclidean distances and model the optimization constraints. The core idea behind this design relies on the capability of neural networks to approximate non-linear transformations to make the Distances Encoder Network learn the spatial transformation that maps initial non-feasible solutions of the constrained unit disk problem into feasible ones. The proposed approach outperforms classical solvers, given fixed comparable computation times, and paves the way to address other optimization problems through a similar strategy.
\end{abstract}

\begin{IEEEkeywords}
embedding, graphs, neural networks, neutral atoms, optimization
\end{IEEEkeywords}

\section{Introduction}\label{sec:intro}
In recent years, quantum computers have garnered more and more interest, as they represent very auspicious tools to accelerate specific computations like material simulations \cite{buluta2009quantum, kandala2017hardware, mcardle2020quantum}, combinatorial optimization \cite{kochenberger2014unconstrained, kadowaki1998quantum, farhi2014quantum} etc. However, we are currently in the noisy intermediate-scale quantum (NISQ) era. Thus practical applications of canonical quantum algorithms (\textit{e.g.}, Shor’s or Grover's algorithms) are still unattainable because of technical limitations such as low qubits count, limited coherence time, and gate fidelity. To overcome these limitations, different approaches emerged, like hybrid quantum-classical algorithms.

In this regard, our paper introduces a novel approach to exploiting neural networks (NN) to optimize neutral atoms' quantum architectures \cite{grimm2000optical}. More precisely, the proposed framework deals with combinatorial optimization problems embedding into the previously mentioned quantum machines. The embedding task from an optimization point of view is equivalent to solving a constrained unit disk graph (CUDG) problem. This optimization problem concerns finding a unit disk graph (UDG) \cite{clark1990unit} (see Def. \ref{def:udg}) realization, given a generic graph. Contextualizing the optimization problem in a real-world quantum computing (QC) application brings additional constraints to a standard UDG problem, thus completing its definition.\\

\begin{definition}[Unit disk graph]\label{def:udg}
Consider $n$ circles, with radius $r$, in the plane. The intersection graph \cite{golumbic2004algorithmic} of these circles is a unit disk graph with $n$ vertexes where each vertex corresponds to a circle centre and the vertexes share an edge only if the corresponding circles intersect. 
\end{definition}

\subsection{Unit disk graphs in quantum applications}

Unit disk graphs have been popularized, in real-world applications, by wireless communication \cite{balasundaram2009optimization, 7579144, li2003algorithmic}. However, with the emergence of quantum technologies, UDGs have also become of interest in the quantum field. In particular, quantum neutral atoms machines \cite{grimm2000optical} rely on Rydberg atoms, positioned on a 2D/3D register, to represent \emph{qubits}. From the interactions between neutral atoms, subject to the action of laser pulses, a spin Hamiltonian (\textit{e.g.}, Ising) is retrieved. The evolution of the spin Hamiltonian is related to qubits (neutral atoms) that, once measured, assume one out of two possible quantum states (\textit{i.e.}, the excited Rydberg state $\vert 1 \rangle$ or the ground state $\vert 0 \rangle$), hence the association with binary optimization variables. This capability of neutral atoms machines enables the mapping of a large set of NP-hard combinatorial optimization problems into the Ising Hamiltonian \cite{lucasIsingFormulationsMany2014}.

In principle, the class of problems that can benefit from this quantum-based solution paradigm \cite{serret2020solving,pasqal_optim} are Quadratic Unconstrained Binary Optimization (QUBO) problems \cite{glover2018tutorial}. They are characterized by a square matrix $Q \in \mathbb{R}^{n\times n}$ and a vector of $n$ binary variables $x \in \{0,1\}^{n}$. The complete definition of QUBO problems is obtained by minimizing the objective functions in the form $x^TQx$. Theoretically, once the qubits are organised in the space to reproduce the desired Hamiltonian, the associated QUBO problem can be solved using Quantum Approximate Optimization Algorithm \cite{serret2020solving,pasqal_optim} or Quantum Adiabatic Algorithm \cite{albash2018adiabatic}.

In neutral atoms' quantum architectures, the blockade effect is one of the main players. It is a threshold-like effect reached when the distances between qubit pairs are shorter than the blockade radius \cite{ciampini2015ultracold}, \textit{i.e.} a critical distance at which the strength of the interactions balances with the Rabi frequency of the laser pulses \cite{picken2018entanglement}. Therefore, the interactions between qubits in the register induce a UDG. In this situation, the halved blockade radius plays the role of the radius $r$, as in definition \ref{def:udg}, and its appropriate value can be set according to the characteristics of the considered quantum device. The qubit positions instead correspond to the circle centres, so when $n$ qubits are placed on the quantum register and are excited through laser pulses, the corresponding $n$-vertexes UDG can be retrieved.

\begin{figure}[ht]%
    \centering
    
    \subfloat[\centering Adjacency matrix]{{\includegraphics[width=0.75\linewidth]{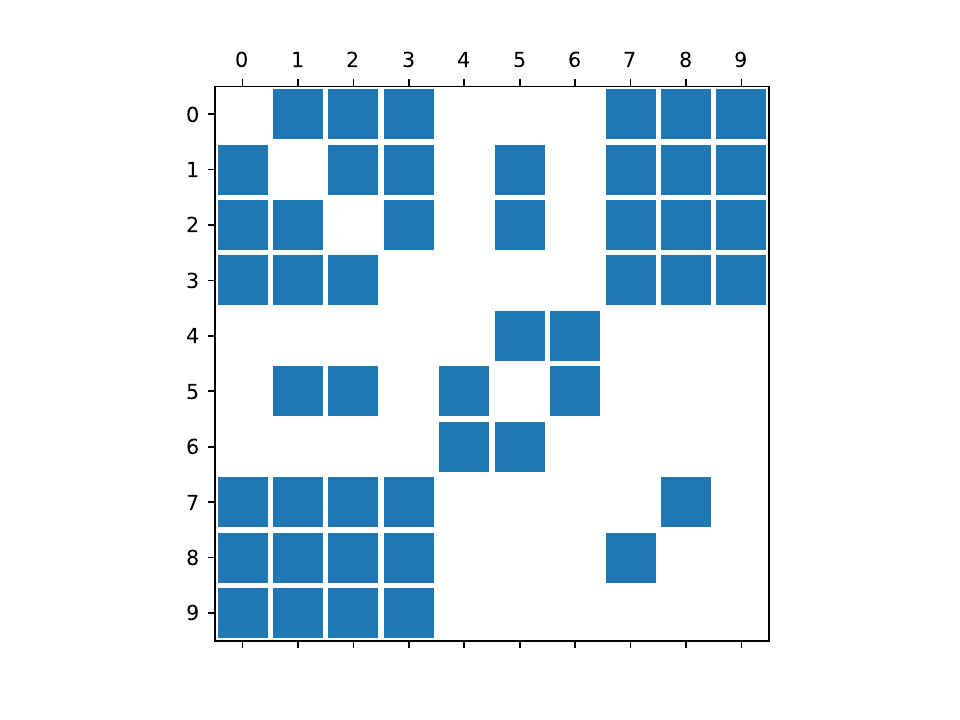} \label{fig:adj_mat} }}\\
    \subfloat[\centering Unfeasible graph embedding]{{\includegraphics[width=0.75\linewidth]{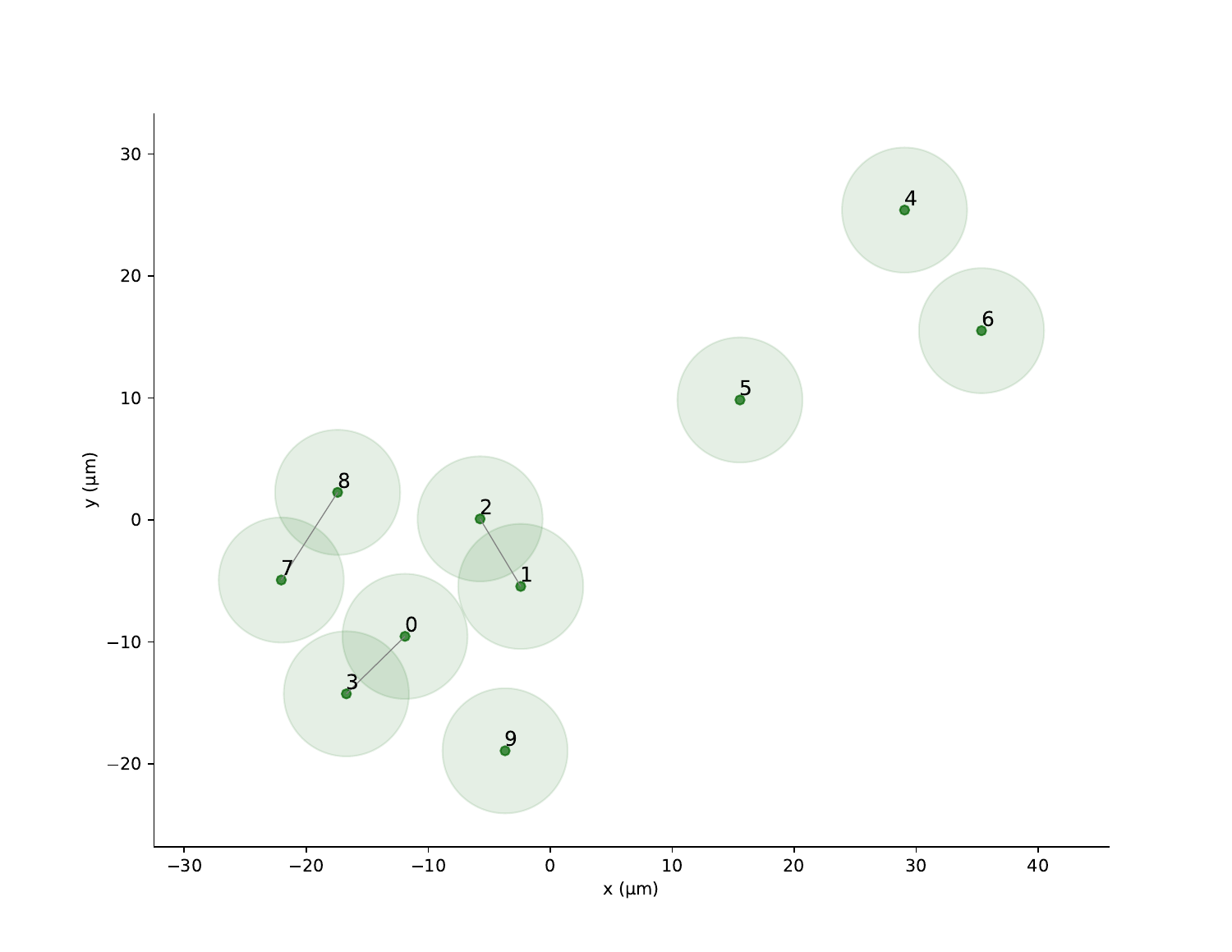} \label{fig:unfeasible} }}\\
    \subfloat[\centering Feasible graph embedding]{{\includegraphics[width=0.75\linewidth]{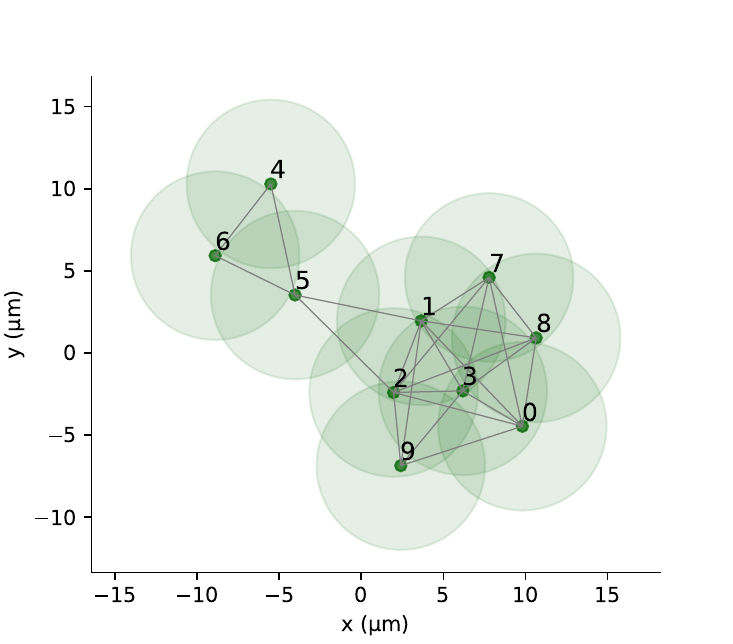} \label{fig:feasible} }}%
    
    \caption{Representation of unit disk graphs in a quantum register. Whenever two circles intersect an edge is generated. The radius of the circles in figs.\ref{fig:unfeasible},\ref{fig:feasible} is the half blockade radius, \textit{i.e.} $r$. The adjacency matrix in fig.\ref{fig:adj_mat} describes the desired edges, it determines the feasibility of the embedding.}
\end{figure}

However, the representation of the spin Hamiltonian for application use cases usually requires a backward approach: the off-diagonal elements of $Q$ describe the connectivity pattern (\textit{i.e.} the adjacency matrix) wanted in the quantum register. Atoms (\textit{i.e.} qubits) are placed in 2D/3D configurations, and their interactions define UDGs, as reported in Fig \ref{fig:unfeasible} and Fig \ref{fig:feasible}. In sum, qubit positions are looked for so that the blockade effect reproduces a UDG that respects the desired connectivity.

Moreover, other requirements come along with the specific quantum device, thus adding constraints to the UDG problem. For the hardware considered in this work, the tweezers governing qubits placement can not place atoms nearer than $D_{min}= 4\ \mu m$, the register can handle atoms placed within a circular area of radius $L=50\ \mu m$, and the greatest value allowed for radius $r$ is estimated at $D_{adj} \approx 10.26\ \mu m$. Finally, it is desirable that the qubits placement not only induces the wanted UDG configuration but also corresponds to a UDG solution that maximizes the \textit{adjacency gap}. That means maximizing the difference between the minimum distance among qubits pairs that are not subject to the blockade effect (vertexes of the UDG not paired by an edge) and the unit disk radius $r$ for interacting qubits (adjacent vertexes in the UDG).

\section{Related work}
Since the embedding on the neutral atoms quantum architecture corresponds to finding CUDG solutions and our methodology relies on neural networks, we investigated literature under two main topics. On one hand, state-of-the-art approaches and results in the context of UDG problem solutions are analyzed by pointing out their limitations concerning our use case. On the other hand, previous work targeted the solution of optimization problems through NNs, so they provide useful insights when attempting similar approaches.

\subsection{Solving the unit disk graph problem}

Solving the CUDG problem is a complex task on different levels. Indeed solving the UDG recognition problem, \textit{i.e.} determining if a given graph has a UDG realization, is NP-hard \cite{breuUnitDiskGraph1998}. Moreover, even retrieving an approximate solution to the UDG problem is impossible in polynomial time unless $P=NP$ \cite{kuhnUnitDiskGraph2004a}. The UDG problems do not become easier to solve for simplified graphs subclasses, such as for outerplanar \cite{syslo1979characterizations} and tree graphs \cite{bhoreUnitDiskRepresentations2021}.

Despite the complexity of the problem, previous works proposed approximation algorithms, especially in the research field of wireless communication, computing virtual coordinates of sensor networks \cite{moscibroda2004virtual, pemmaraju2011good, rusterholz2003approximation, zhou2010practical}. Unfortunately, for the quantum application targeted by this paper, the proposed approximations are not feasible. Furthermore, the additional requirements deriving from the quantum device introduced more non-convex constraints. Thus, the standard approach for the CUDG problem solution requires defining the non-convex programming model and trying to solve it with classical solvers.  On this side, different formulations for the CUDG problem could be designed, as in \cite{pasqal_optim}, and according to the class of the programming model, suitable solvers could be exploited. Mixed-Integer Quadratic Constrained Programming can be solved with \textit{Gurobi} \cite{bixby2007gurobi} (which nevertheless performed poorly for the formulation proposed in Sec. \ref{opt_problem}), and Non-linear Programming can be approached with \textit{Ipopt} \cite{wachter2006implementation}. \textit{Ipopt} solver, due to preliminary better results, has been chosen as a classical solver for comparisons. 

\subsection{Applying neural networks to solve optimization problems}
Previous studies investigated the application of neural networks to solve optimization problems. In \cite{fischetti2018deep}, binary Mixed Integer Linear Programs are handled through NNs by devising proper architectures: \textit{ReLu} activation functions implement the binary variables, whilst continuous variables are directly represented by the output value of each unit. The methodology is applied to \textit{feature visualization} and \textit{adversarial machine learning} tasks. A similar approach was presented by Amos and Kolter \cite{amos2017optnet} to solve Quadratic Programs. They provide examples of learning Sudoku problems.

Beyond that, more specific applications of neural networks for optimization are present in the literature. Chandrasekhar and Suresh exploited the weights and biases of NNs to parametrize a density function for topology optimization \cite{chandrasekhar2021tounn}. In \cite{chandrasekhar2021multi}, they extended their work to deal with multi-material topologies. In \cite{ardon2022reinforcement}, Reinforcement Learning enhances the solution of Capacitated Vehicle Routing Problems, providing a trained policy to solve unseen instances.

Graphs-based optimization (\textit{e.g.}, Vertex Cover, Maximum Independent Set problems) is instead the subject of \cite{abe2019solving, mccarty2021nn}. The proposed methodologies consider Convolutional and Graph NNs for their optimization purposes.

\section{Method}\label{sec:methodology}
\subsection{The constrained unit disk graph problem}\label{opt_problem}
The QC use case, as described in section \ref{sec:intro}, requires a specific formulation of the CUDG optimization problem. Here, we propose a programming model that considers both the unit disk properties and the quantum hardware constraints.

Before diving into a detailed description of the programming model, some remarks are needed. The proposed formulation exploits binary variables to enforce the constraints: a theoretically equivalent model could be designed with only continuous variables. However, binary variables do not affect the convexity of the problem; the formulation with only continuous variables is nonetheless non-convex. Furthermore, these binary variables are needed to deal with the State-of-the-Art solver \textit{Ipopt} \cite{wachter2006implementation} through the \textit{Pyomo}\footnote{https://pyomo.readthedocs.io} Python library. This solver allows constraint violations through tolerance parameter settings, but even the most stringent tolerance does not prevent numerical issues (\textit{e.g.}, numerical cancellation and errors inherent in floating-point arithmetic). Thus it may lead to unfeasible solutions when the feasibility is not enforced through binary variables, explicitly reflecting constraint violations in the objective function. Beware that in this specific quantum application, the requirements correspond to hard constraints, \textit{i.e.} the quantum hardware cannot deal with qubits positions that are approximately feasible.

To mathematically define the constrained unit disk graph (CUDG) problem, we introduce the following notation.\\
$\mathcal{G}(\mathcal{V}, \mathcal{E})$ represents the undirected graph to embed, with $\mathcal{V}$ as the set of vertexes and $\mathcal{E}$ as the set of undirected edges.
The number of vertexes $|\mathcal{V}|$ will be denoted as $n$, and the vertexes' labels will be indexed starting from $0$; $\mathcal{P}$ is the set of all unordered pairs in $\mathcal{V}$, so $|\mathcal{P}|=\frac{n(n-1)}{2}$.
The square matrix $A$ of size $n \times n$ is the adjacency matrix of $\mathcal{G}$.\\
The parameterization of the CUDG problems' instances is determined by the positive constants $D_{min}, D_{adj}$ and $L$ that define the feasibility domain. Respectively, they represent the minimum allowed distance between vertex pairs, the maximum allowed distance between adjacent vertexes, and the maximum radius of the circle/sphere inscribing the graph embedding.
Finally, the embedding dimensionality will be defined as $N \in \{2,3\}$.

Regarding the CUDG programming model, the coordinates of the vertexes are represented by $N$-dimensional vectors $\overrightarrow{p}_i, \quad \forall i \in \mathcal{V}$. These are continuous variables in the square/cubic domain of side $2L$ (see eq. \eqref{cons:coords}).
The maximum distance between adjacent pairs is modelled through the continuous variable $d_{adj}$, eq. \eqref{domain:adj}. The minimum distance between not adjacent pairs is defined by another continuous variable $d_{\overline{adj}}$, eq. \eqref{domain:notadj}.
At last, the binary variables $\delta_{ij}$, $\forall \{i,j\} \in \mathcal{P}$, defined as follows, model the feasibility of the solution.
\begin{equation}
\delta_{ij} := 
\begin{cases}
	1 & $pair distance is unfeasible $\\
	0 & $pair distance is feasible $
\end{cases}
\end{equation}

 It is relevant to notice that the feasibility conditions are modelled accordingly to the adjacency pattern described by $A$ to account for the unit disk graph property and constraints: adjacent vertexes' feasible distances are in the range $[D_{min}, D_{adj}]$ (see Eqs. \eqref{cons:adj_min}, \eqref{cons:adj_max}), not adjacent vertexes should have pair distances in the range $[D_{adj}+\epsilon, 2L]$ (see Eqs. \eqref{cons:notadj_min}, \eqref{cons:notadj_max}), with a small value $\epsilon$ to avoid a strict inequality formulation that is not allowed by the \textit{Pyomo} library.
 Constraint \eqref{cons:maxmax_adj}, combined with the objective function \eqref{obj}, enforces the adjacent vertexes to be as close as possible, whereas constraint \eqref{cons:minmin_notadj} enhances the distances between not adjacent vertexes to be the greatest as possible.\\
 The overall CUDG programming model is shown below. Notice that the constraints on the distances are defined taking into account squared Euclidean distances and that the penalty constant $2L-D_{min}+\iota,\ \iota>0$, associated with each binary variable in the objective function, favours feasibility over the increasing of the \textit{adjacency gap}, $d_{adj} - d_{\overline{adj}}$.
 
{\footnotesize
\begin{mini!}|s|
{\overrightarrow{p}, \delta,  d}{(2L-D_{min}+\iota)\sum_{\{i,j\} \in \mathcal{P}} \delta_{ij} + d_{adj} - d_{\overline{adj}} } {}{}\label{obj}
\addConstraint{||\overrightarrow{p}_i-\overrightarrow{p}_j||_2^2}{\leq D_{adj}^2 + (8L^2-D_{adj}^2)\delta_{ij} \quad & (i,j) \in \mathcal{E}} \label{cons:adj_max}
\addConstraint{||\overrightarrow{p}_i-\overrightarrow{p}_j||_2^2}{\geq (1-\delta_{ij})D_{min}^2 \quad & (i,j) \in \mathcal{E}} \label{cons:adj_min}
\addConstraint{||\overrightarrow{p}_i-\overrightarrow{p}_j||_2^2}{\leq d_{adj}^2 \quad & (i,j) \in \mathcal{E}} \label{cons:maxmax_adj}
\addConstraint{||\overrightarrow{p}_i-\overrightarrow{p}_j||_2^2}{\leq 4L^2 + 4L^2\delta_{ij} \quad & (i,j) \notin \mathcal{E}}\label{cons:notadj_max}
\addConstraint{||\overrightarrow{p}_i-\overrightarrow{p}_j||_2^2}{\geq (1-\delta_{ij})(D_{adj}+\epsilon)^2 & (i,j) \notin \mathcal{E}}\label{cons:notadj_min}
\addConstraint{||\overrightarrow{p}_i-\overrightarrow{p}_j||_2^2}{\geq d_{\overline{adj}}^2 & (i,j) \notin \mathcal{E}} \label{cons:minmin_notadj}
\addConstraint{\overrightarrow{p}}{\in [-L,+L]^{n \times N}} \label{cons:coords}
\addConstraint{d_{adj}}{\in [D_{min},D_{adj}]} \label{domain:adj}
\addConstraint{d_{\overline{adj}}}{\in [D_{adj}+\epsilon,2L]} \label{domain:notadj}
\addConstraint{\delta}{\in \{0,1\}^{|\mathcal{P}|}}
\end{mini!}\par}

\subsection{Generating the dataset: basic requirements}\label{sec:dataset_creation}
The programming model presented in section \ref{opt_problem} is solved by comparing two solvers: the \textit{Ipopt} solver and the novel neural-enhanced optimization framework detailed in Sec. \ref{nn_opt}.
To fairly compare these solvers, a test dataset has been created: it consists of $200$ graph instances; in particular, there are $20$ samples for each value of $n \in \{10,20,\ldots,100\}$.

It is worth mentioning that the generation of the dataset takes into account some necessary conditions regarding the feasibility of the solution to the CUDG problems.
The identified necessary conditions  follow directly from \emph{Thue's Theorem}, which states that \textit{regular hexagonal packing is the densest circle packing in the plane} \cite{chang2010simple}.
As follows, necessary conditions have been defined for the 2-dimensional case, which corresponds to the most quantum-application-ready setting. Further generalization to the 3-dimensional case or more sophisticated conditions will be the subject of future work.

\begin{figure}[ht]
\centerline{\includegraphics[width=1.0\columnwidth]{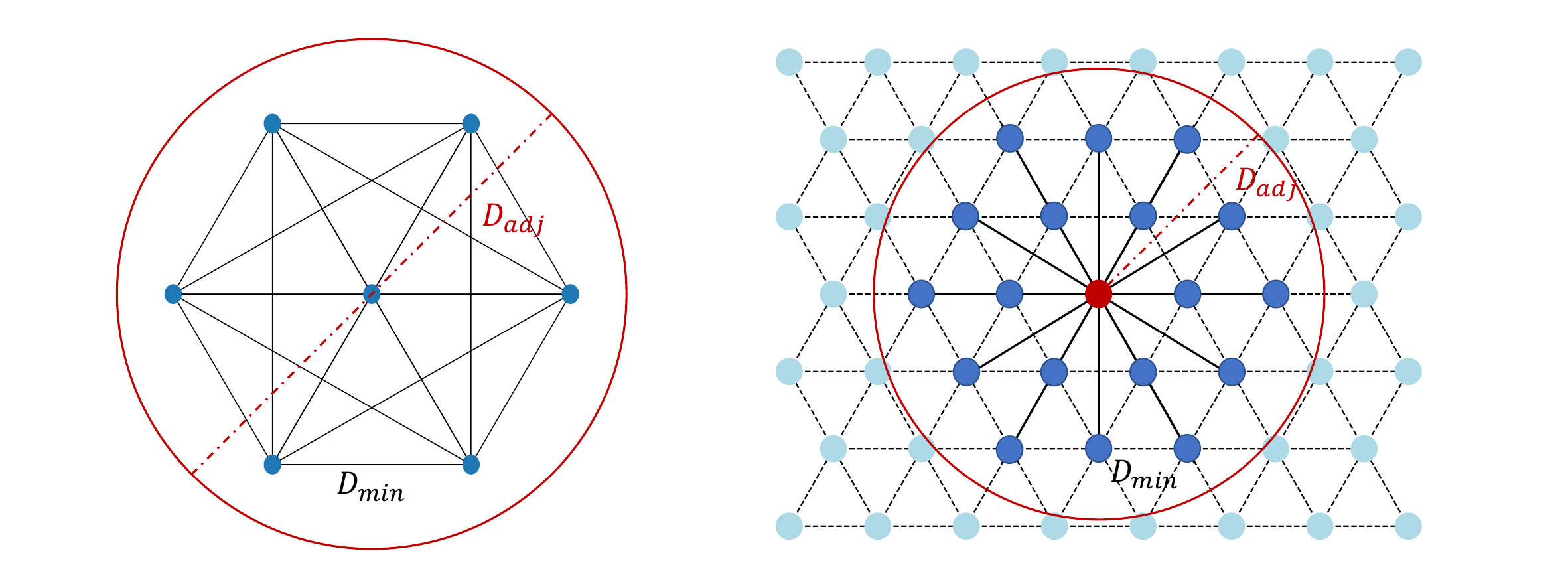}}
\caption{Representations of a $K_7$, clique with $7$ vertexes, (on the left) and of a graph with maximum degree $\Delta=18$ (on the right) feasible embeddings, for a CUDG problem with $D_{min}=4\ \mu m$, $L=50\ \mu m$ and $D_{adj}=10.26\ \mu m$.}
\label{fig:triang_emb}
\end{figure}

Concerning the real-world application, the domain parameters values are $D_{min}=4\ \mu m$ and $D_{adj} \approx 10.26\ \mu m$. So, in the densest packing embedding, the hexagon side is $4\ \mu m$, and adjacent vertexes should lie within a $\approx 10.26\ \mu m$ distance.
Thus, to properly embed a complete graph with $M$ vertexes, $K_M$, all its vertexes should have pair distances $\leq D_{adj}$ (see Fig. \ref{fig:triang_emb} on the left), hence property \ref{property:max_clique}.\\

\begin{property}[Maximum clique property]\label{property:max_clique}
Consider the CUDG problem formulated in section \ref{opt_problem}. Let $D_{min}=4\ \mu m$, $D_{adj} = 10.26\ \mu m$, $L=50\ \mu m$, $\mathcal{G}(\mathcal{V},\mathcal{E})$ the undirected graph that defines the CUDG problem and $M$ the number of vertexes in the maximum-sized complete graph (or clique) of $\mathcal{G}$, then a feasible unit disk graph solution can be obtained only if $M\leq 7$.
\end{property}
\vspace{\baselineskip}
Moreover, still considering the same regular hexagonal packing combined with the unit disk graph definition, property \ref{property:max_degree} follows.

\begin{property}[Maximum degree property]\label{property:max_degree}
Consider the CUDG problem formulated in section \ref{opt_problem}. Let $D_{min}=4\ \mu m$, $D_{adj} = 10.26\ \mu m$, $L=50\ \mu m$, $\mathcal{G}(\mathcal{V},\mathcal{E})$ the undirected graph that defines the CUDG problem and $\Delta$ the maximum degree for vertexes in $\mathcal{G}$, then a feasible unit disk graph solution can be obtained only if $\Delta \leq 18$.
\end{property} 

Fig. \ref{fig:triang_emb}, on the right, shows a feasible embedding for a subgraph with $\Delta=18$. All the vertexes lying within the circular red area, with radius $D_{adj}=10.26\ \mu m$, are neighbours of the red-colored vertex, which among the highlighted vertexes is the one with the highest degree $\Delta=18$. However, the represented embedding constraints the connectivity of neighbouring vertexes. Thus it is not a feasible embedding for all arbitrary graphs with $\Delta=18$.

So, since properties \ref{property:max_clique} and \ref{property:max_degree} define the necessary conditions for the CUDG to have a feasible solution, they govern the dataset generation. To have more chance and get feasible embeddings, the graphs instances are created by randomly setting initial coordinates $\overrightarrow{c}_i$, $\forall i \in \mathcal{V}$ in a square domain with side $l$. Then, edges are defined following a unit disk graph approach, considering a threshold distance $d$: all vertexes that fall within distance $d$ are paired by edges. No constraints on minimum feasible distance between vertexes or tight relationship between values $d$ and $l$ are considered. So the CUDG problem, as defined in section \ref{opt_problem}, is not trivially solved by scaling the initial domain. The parameters $l$ and $d$ increment along with the number of vertexes.
The dataset creation is performed iteratively until all the desired samples are obtained. In the specific case, the overall dataset, accounting for $200$ graphs, was computed in $\approx 2.30\ min$.

Therefore, in generating the dataset, all the graphs' instances were required to satisfy the following conditions:
\begin{itemize}
    \item size of the maximum estimated\footnote{Finding the maximum clique is an NP-hard problem, which solution was not targeted in this context, the \textit{NetworkX} maximum clique approximation algorithm was exploited at this scope \cite{boppana1992approximating}.} clique $\leq 7$;
    \item maximum degree $\leq 18$;
    \item all vertexes of the graph belong to the same connected component.
\end{itemize}

\subsection{Neural-enhanced optimization framework}\label{nn_opt}

As classical solvers’ performance significantly decreases with CUDG problem dimensionality, we designed a novel methodology to enable solving more CUDG instances.
This approach exploits neural networks’ capability to approximate non-linear functions to make a modified autoencoder learn the spatial transformation that maps an initial unfeasible solution of the CUDG into a feasible one, possibly satisfying both the unit disk graph property and maximizing the \textit{adjacency gap}.\\
The overall model, named \textit{Distance Encoder Network (DEN)}, takes as input an initial guess on the coordinates $\overrightarrow{p}_i$, learns new feasible coordinates $\overrightarrow{p}_i^f$  in a specific hidden layer, \textit{i.e.} the \textit{coordinates' layer}, and produces as output the squared pair distances $||\overrightarrow{p_i}^f-\overrightarrow{p_j}^f||_2^2,\ \forall \{i,j\} \in \mathcal{P}$.
Then, the CUDG problems' constraints and the objective function are handled through a custom loss function, named \textit{Embedding Loss Function (ELF)}.

\begin{figure*}[ht]
\centerline{\includegraphics[width=\linewidth]{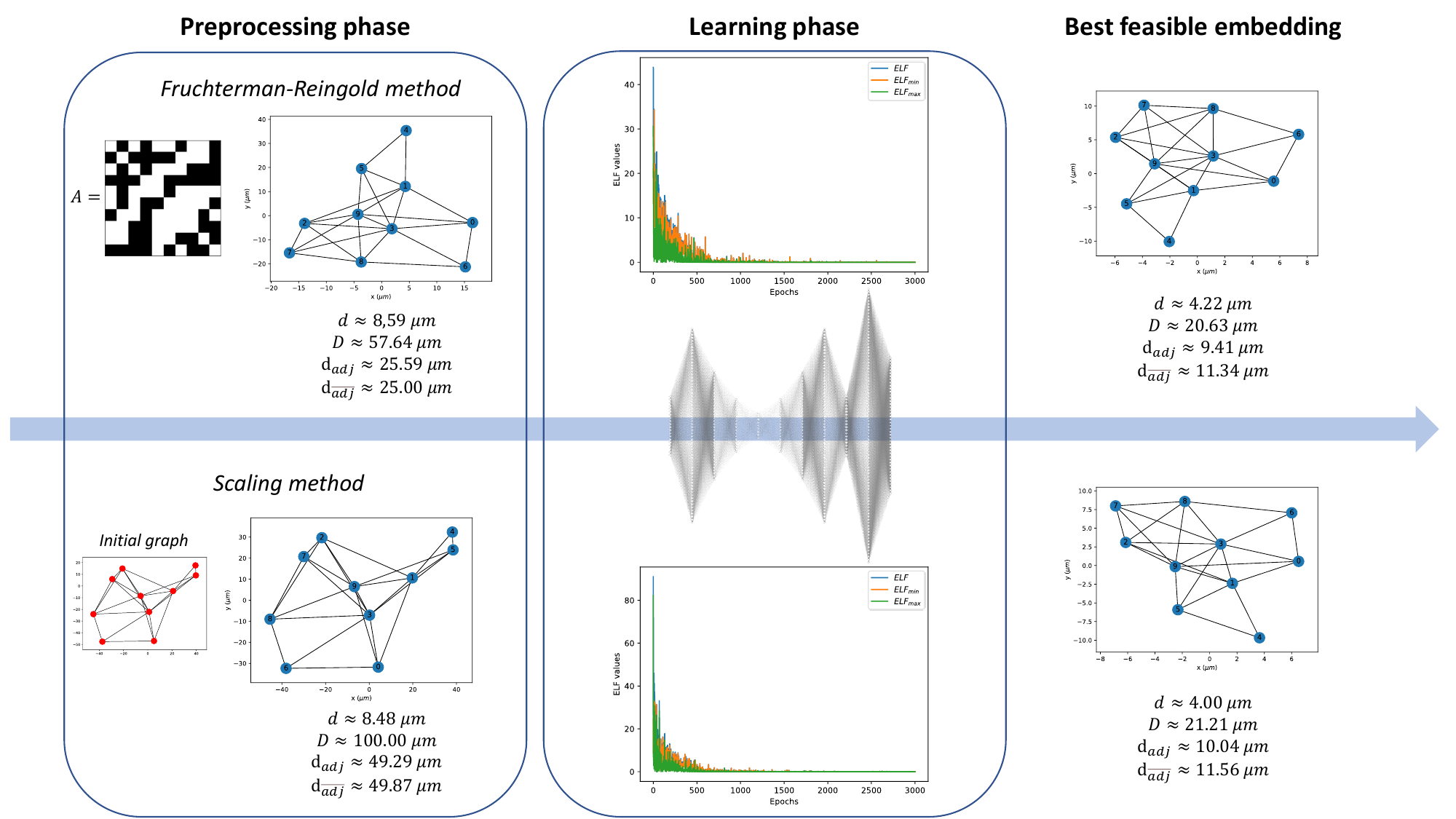}}
\caption{The optimization framework for the constrained unit disk graph problem: two different approaches are considered for the position initializations in the \textbf{preprocessing phase}.
Both the \textit{Scaling} and the \textit{Fruchterman-Reingold methods} do not provide initial feasible embeddings.
Then, during the \textbf{learning phase}, the \textit{DEN} model has $3000$ epochs available for learning a proper spatial transformation.
Along with the model training, the best feasible embeddings are updated according to the \textit{adjacency gap} maximization goal.
The final step returns the best embedding found so far.}
\label{fig:opt_framework}
\end{figure*}

So, the overall neural-enhanced optimization framework manages one graph instance $\mathcal{G}(\mathcal{V}, \mathcal{E})$ at a time. As represented in Figure \ref{fig:opt_framework}, it consists of two main parts: the \textbf{preprocessing phase} provides $\overrightarrow{p}_i$, $\forall i \in \mathcal{V}$, and then the \textbf{learning phase} performs the actual optimization to retrieve $\overrightarrow{p}_i^f$, $\forall i \in \mathcal{V}$ for a feasible embedding configuration.

\vspace{\baselineskip}
\subsubsection{The preprocessing phase: the vertex coordinates initialization}
Since the proposed optimization framework relies on learning spatial transformations, initial coordinates $\overrightarrow{p}_i$, $\forall i \in \mathcal{V}$ should be provided. Hence, a preprocessing phase performs the computation of these initial positions. This preliminary phase aims to support the convergence of the optimization algorithm, \textit{i.e.} providing $\overrightarrow{p}_i$, $\forall i \in \mathcal{V}$ that roughly satisfy some of the constraints. Thus, the \textit{DEN} model would converge in fewer iterations (epochs) than when starting from random initialization.
Two approaches for $\overrightarrow{p}_i$, $\forall i \in \mathcal{V}$ initialization have been investigated.
\begin{itemize}
  \item \textit{Scaling method}: all the graphs in the dataset come along with vertex coordinates $\overrightarrow{c}_i$, $\forall i \in \mathcal{V}$.
These coordinates can be scaled to the domain of interest to respect at least one set of distance constraints.
The choice was made to scale $\overrightarrow{c}_i$, $\forall i \in \mathcal{V}$ to a circle with radius $L$, thus retrieving $\overrightarrow{p}_i$, $\forall i \in \mathcal{V}$ initial positions that automatically satisfy constraints \eqref{cons:notadj_max}.
If the target is a 3-dimensional embedding, \textit{i.e.} $N=3$, then all z-coordinates are initialized to $0$.
\item \textit{Fruchterman-Reingold method}: this force-directed layout algorithm \cite{fruchterman1991graph} does not require initial coordinates to be performed, and it models attractive and repulsive forces between vertex pairs according to the adjacency pattern. The chosen Fruchterman-Reingold algorithm implementation is available in the \textit{NetworkX}\footnote{https://networkx.org} Python library, where repulsive forces intervene on all vertex pairs with module $k^2/||\overrightarrow{p}_i-\overrightarrow{p}_j||_2^2$, whilst attractive forces intervene only on adjacent pairs and have module $||\overrightarrow{p}_i-\overrightarrow{p}_j||_2/k$. Here the parameter $k$ determines the equilibrium distance at which the two forces balance for adjacent pairs \cite{hagberg2008exploring}, so in this specific setting, it assumes a value of  $7\ \mu m \in [D_{min},D_{adj}]$.
This method does not guarantee some constraints satisfaction through  $\overrightarrow{p}_i$, $\forall i \in \mathcal{V}$ initialization; though, it allows to deal with graph instances that do not have any initial coordinates to start from, a valid assumption for most of the graphs in UDG-related applications. It handles both 2D and 3D coordinate vectors. This is an iterative method, which tends to converge to a solution in a short time. Thus, it was restrained to $1000$ iterations.
\end{itemize}
\vspace{\baselineskip}

\subsubsection{The learning phase: pursuing the feasible embedding}

After the \textbf{preprocessing phase}, the core of the optimization framework takes place. It consists of the \textit{DEN} model's training.
Even if the training algorithm follows the typical approach to training neural networks (forward step to compute outputs and gradients, and backward step to update the network's weights), it has a different meaning. The \textit{DEN} model is supposed to learn a proper spatial transformation that maps an initial not feasible solution of the CUDG problem into a feasible one, still targeting the maximization of the \textit{adjacency gap}. So, the training of each graph sample is independent of the others. Hence, the network architecture is defined according to the problem dimensionality, \textit{i.e.} $n$ and $N$, and to the desired adjacency pattern, determined by $A$. To set up a fair comparison among graph samples, for each instance of the CUDG problem, a maximum number of epochs, \textit{i.e.} \textit{DEN} model trials to find a solution, is fixed, and it is denoted by $E$.\\
The \textit{DEN} model architecture includes dropout layers for regularization purposes \cite{wager2013dropout}. Therefore, at each epoch, a \textit{training step} and a subsequent \textit{inference step} are performed: during the \textit{training step}, the dropout works by randomly and temporarily deleting neurons in the hidden layers, then the \textit{ELF} is computed, and \textit{DEN} weights are updated according to \textit{AdamW} optimizer with learning rate defined by the hyperparameter $lr$ \cite{loshchilov2017decoupled}, after that, in the \textit{inference step} no dropout takes place, and the embedding configurations is computed without further contributing to weights' update.\\ In the \textit{ELF} definition, the parameter $\alpha$ represents the \textit{adjacency gap}, so it would not be directly optimized if an outer optimization loop were not performed.
In the proposed solution, $\alpha$ is initialized to $\epsilon$, and each time the \textit{DEN} model finds a feasible solution, if the new solution corresponds to an increment in the \textit{adjacency gap}, the parameter $\alpha$ assumes the value of the best \textit{adjacency gap}.\\ This overall procedure, concerning both the feasible embedding retrieval and the \textit{adjacency gap} maximization, is the \textbf{learning phase}.\\

\textbf{\textit{The Distance Encoder Network architecture}}: the \textit{DEN} model architecture comes from a modification of a typical autoencoder network. It consists of two parts, which are respectively the \textit{trainable autoencoder} component and the \textit{fixed-weight distance calculator}.\\
The \textit{trainable autoencoder} has the architecture described in table \ref{tab:autoencoder}: the input layer accounts for all the initial coordinates components, $\overrightarrow{p}_i$, $\forall i \in \mathcal{V}$ and the output layer generates the transformed coordinates, $\overrightarrow{p_i}^f$, $\forall i \in \mathcal{V}$. In particular, these positions are flattened into $1$-dimensional vectors, the input vector $I$ and the output vector $O$, such that

\begin{equation}
    I(k) = 
    \begin{cases}
    	\overrightarrow{p}_{k}^{(x)} & 0 \leq k < n\\
        \overrightarrow{p}_{k-n}^{(y)} & n \leq k < 2n\\
        \overrightarrow{p}_{k-2n}^{(z)} & 2n \leq k < 3n \quad $and$ \quad N=3\\
    \end{cases}
\end{equation}

\begin{equation}
O(k) = 
\begin{cases}
	\overrightarrow{p_{k}^{f}}^{(x)} & 0 \leq k < n\\
    \overrightarrow{p^{f}}_{k-n}^{(y)} & n \leq k < 2n\\
    \overrightarrow{p^{f}}_{k-2n}^{(z)} & 2n \leq k < 3n \quad $and$ \quad N=3\\
\end{cases}
\end{equation}

In the \textit{trainable autoencoder}, all the fully connected layers allow for the contribution of a bias node and they are equipped with the dropout functionality. This latter is parameterized through the dropout probability hyperparameter $p_{drop}$.\\ 
Finally, the activation function in the last fully connected layer allows generating coordinates in the square (or cubic, when $N=3$) domain of side $2L$.

\begin{table*}[htbp]
\caption{\textit{Trainable autoencoder} component architecture for a graph with $n$ vertexes and targeting an embedding in $N$ dimensions.}
    \centering
    \begin{tabular}{ccccc}
    \toprule
        &  Layer type & Input size & Ouput size & Activation function\\
        Encoder    & & & &        \\
                           & Fully connected layer & $n \times N $ & $64$ & $ReLu$ \\
                           & Fully connected layer & $64 $ & $36$ & $ReLu$ \\
                           & Fully connected layer & $36 $ & $18$ & $ReLu$ \\
                           & Fully connected layer & $18 $ & $9$ & $ReLu$ \\
        \midrule
        Decoder           & & & &        \\
                          & Fully connected layer & $9 $ & $18$ & $ReLu$ \\
                          & Fully connected layer & $18$ & $36$ & $ReLu$ \\
                          & Fully connected layer & $36$ & $64$ & $ReLu$ \\
                          & Fully connected layer & $64$ & $n \times N $ & $L\times Tanh$ \\
      \bottomrule
    \end{tabular}
    
    \label{tab:autoencoder}
\end{table*}

The \textit{fixed-weight distance calculator} is the second component of the \textit{DEN} model. It computes squared pair distances, $||\overrightarrow{p_i}^f-\overrightarrow{p_j}^f||_2^2,\ \forall \{i,j\} \in \mathcal{P}$, thus making the \textit{trainable autoencoder} aware that the values contained in $O$ represent Cartesian coordinates whilst providing the proper input to the loss function.
To perform this task, the weights of the \textit{distances encoder} are not subject to the training procedure, differently from the \textit{trainable autoencoder}, and the computation of the squared distances is accomplished through the fully connected \textit{difference layer} (input size = $n \times N$, output size = $N \times \binom{n}{2}$, no bias node) followed by a $Square$ activation function and subsequently through the fully connected \textit{sum layer} (input size = $N \times \binom{n}{2}$, output size = $\binom{n}{2}$, no bias node).
In particular, the \textit{difference layer}'s weights assume values $\pm 1, 0$, such that the node values $u$ in this layer, before the activation function is applied, are 

{\small
\begin{align}
    u_{i(n-1)-\binom{i}{2}+j-i-1} &= \overrightarrow{p_{i}^{f}}^{(x)}-\overrightarrow{p_{j}^{f}}^{(x)} \quad \forall \{i,j\} \in \mathcal{P} \\
    u_{(n-1)(\frac{n}{2} + i)-\binom{i}{2}+j-i-1} &= \overrightarrow{p_{i}^{f}}^{(y)}-\overrightarrow{p_{j}^{f}}^{(y)} \quad \forall \{i,j\} \in \mathcal{P}\\
    u_{(n-1)(n + i)-\binom{i}{2}+j-i-1} &=\overrightarrow{p_{i}^{f}}^{(z)}-\overrightarrow{p_{j}^{f}}^{(z)} \quad \forall \{i,j\} \in \mathcal{P},\ N=3
\end{align}\par}

Finally, the weights in the \textit{sum layer} are fixed to values $0,+1$, such that the output values $v$ of the \textit{DEN} models are the squared pair distances in lexicographic order:

{\small
\begin{equation}
v_k = 
    \begin{cases}
    u^{2}_k + u^{2}_{\binom{n}{2}+k} & k \in \Big\{0,1,\ldots, \binom{n}{2}-1\Big\},\ N=2\\
    u^{2}_k + u^{2}_{\binom{n}{2}+k} + u^{2}_{2\binom{n}{2}+k} & k \in \Big\{0,1,\ldots, \binom{n}{2}-1\Big\},\ N=3
\end{cases}
\end{equation}\par}
\vspace{\baselineskip}

\textbf{\textit{The Embedding Loss Function}}: starting from the output of the \textit{DEN} model, $v=||\overrightarrow{p_i}^f-\overrightarrow{p_j}^f||_2^2,\ \forall \{i,j\} \in \mathcal{P}$, the embedding loss function (\textit{ELF}) is defined to address the feasibility constraints. In particular, it handles separately \eqref{cons:adj_min} \eqref{cons:notadj_min} and \eqref{cons:adj_max} \eqref{cons:notadj_max}. They respectively define lower and upper bounds on feasible distances. So, the $\geq$-based constraints are reflected in the loss $ELF_{min}$, whilst the $\leq$-based are modelled through the loss $ELF_{max}$. It is worth mentioning that this modelling approach for inequalities constraints can be exploited beyond the specific CUDG problem.\\
The $ELF_{min}$ and $ELF_{max}$ definitions exploit the \textit{Margin Ranking loss function}:

{\small
\begin{equation}
    MarginRanking(v,v^t,m) = avg (max (0, -m(v-v^t)))
\end{equation}\par}

where $m$ is a vector that determines if we are modelling $\geq$ or $\leq$ inequalities constraints, and the target squared distances $v^t$ are defined according to the adjacency pattern. More precisely, for the $ELF_{min}$ computation: $m=\mathbb{I}_{|\mathcal{P}|}$, and

{\small
\begin{equation}
    v^t_{i(n-1)-\binom{i}{2}+j-i-1}=
    \begin{cases}
    D_{min}^{2}& A_{i,j}=1\\
    (D_{adj}+\alpha)^2 & A_{i,j}=0
    \end{cases}
    \qquad \forall \{i,j\} \in \mathcal{P}
\end{equation}\par}
Whereas, for the $ELF_{max}$, $m=-\mathbb{I}_{|\mathcal{P}|}$, and

{\small
\begin{equation}
    v^t_{i(n-1)-\binom{i}{2}+j-i-1}=
    \begin{cases}
    D_{adj}^2 & A_{i,j}=1\\
    4L^2 & A_{i,j}=0
    \end{cases}
    \qquad \forall \{i,j\} \in \mathcal{P}
\end{equation}\par}

Finally, the overall loss function accounts for both contributions, so
\begin{equation}
    ELF(v) = ELF_{min}(v)+ELF_{max}(v)
\end{equation}

\section{Experiments}
To test the effectiveness of the \textit{DEN}-based optimization framework, $200$ graph samples have been considered to define the corresponding CUDG problem instances. The dataset consists of $20$ graphs for each value of $n \in \{10, 20, \ldots, 100\}$.\\
The feasible domain has been parametrized according to the characteristics of the quantum hardware of interest, so $D_{min}=4\ \mu m$, $D_{adj} \approx 10.26 \ \mu m$, $L=50\ \mu m$. The parameters $\iota$ and $\epsilon$ have been set respectively to $1$ and $0.1$. Each \textit{DEN} model \textbf{learning phase} was allowed to perform $E=3000$  epochs. The learning rate and the dropout probability hyper-parameters were tested with values $lr \in \{0.01, 0.001, 0.0001\}$ and $p_{drop}\in \{0.3, 0.5, 0.7\}$ and combined with the two initialization methods, such as to obtain $18$ different trials for retrieving a CUDG solution for each graph in the dataset.\\
All the trials were run on an IBM Power9-based cluster, with 32 cores/node and 256 GB/node. The nodes in the cluster are also equipped with 4 x NVIDIA Volta V100 GPUs each. However, since the mini-batch size is $1$, \textit{i.e.} just one graph sample is considered at each epoch, training on GPUs does not provide a significant acceleration. Whereas the greater availability of CPUs allowed us to exploit better the trials parallelism. In particular, each trial uses 8 physical cores of a node. The overall experiment, comprehensive of the classical solver comparison, required $\approx 5000$ core hours.

\begin{figure}[ht!]%
    \centering
    \subfloat[\centering Computational time of the \textbf{learning phase} at the varying of the $|\mathcal{V}|=n$]{{\includegraphics[width=\linewidth]{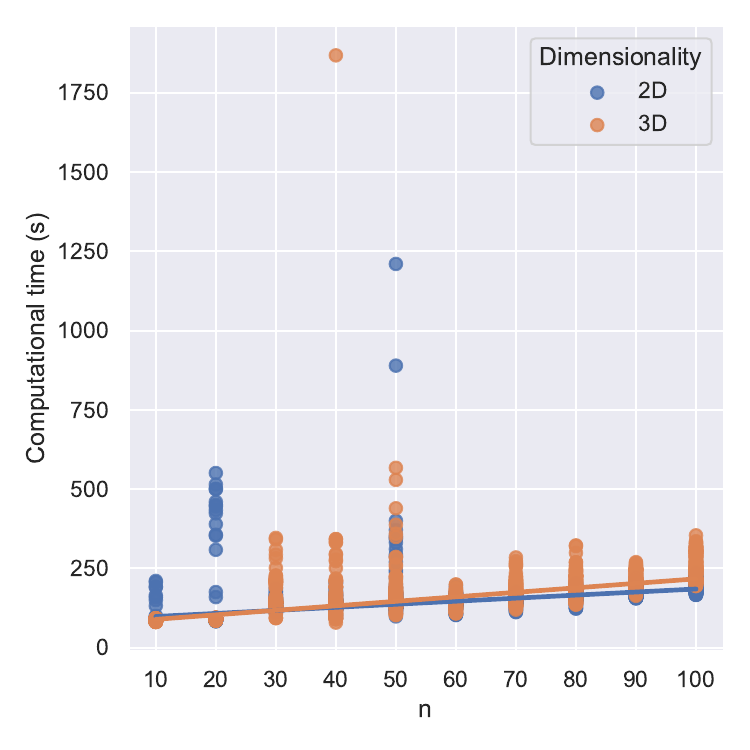} \label{fig:comp_time} }}\\
    \subfloat[\centering Epochs at which the first feasible embedding was found during the \textbf{learning phase}]{{\includegraphics[width=\linewidth]{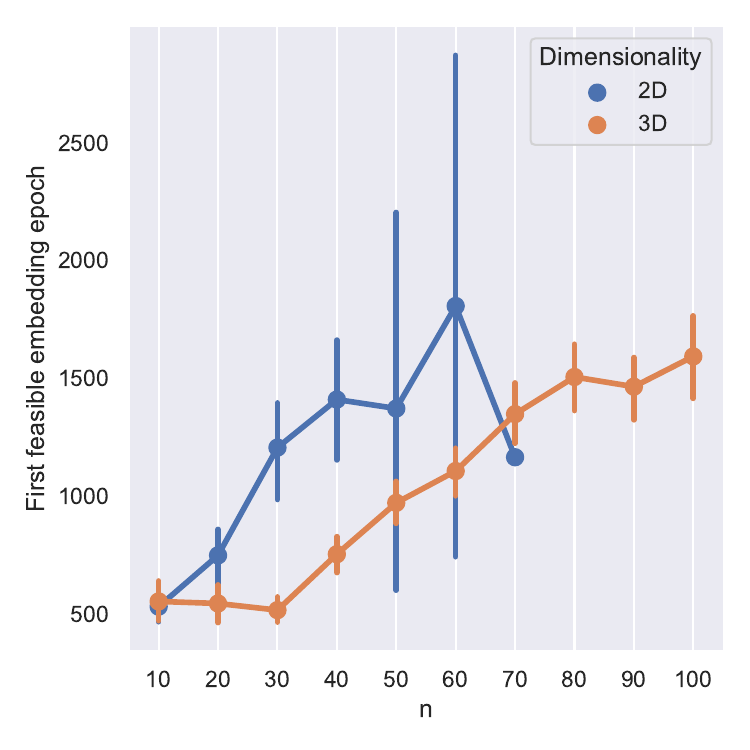} \label{fig:ffepochs} }}%
    \caption{The average computational times for the \textit{DEN} models training scale linearly with $n$, and the increment in the dimensionality does not seem to impact significantly on the training duration (Fig. \ref{fig:comp_time}). Whereas the impact of an increased dimensionality, $N$, is much more evident in Fig. \ref{fig:ffepochs}, here the \textit{DEN} model finds more easily a feasible embedding when $N=3$ than in the case $N=2$, as can be noticed, feasible embeddings are retrieved up to $n=100$ and generally the first feasible embedding is found earlier along the epochs.}%
\end{figure}

Figure \ref{fig:comp_time} shows the variability in computational time to perform each one of the $3600=200\times18$ trials (the dataset consists of $200$ graphs and the combinations in the hyperparameters search are $18$ ) for both the embedding dimensionality $N=2$, $N=3$, grouped by the number of vertexes $n$.
For each value of $n$ and $N$, an average computational time, $T_{n,N}$, is retrieved. Then to perform a comparison with the \textit{Ipopt} solver, a multi-start classical optimization takes place, with the number of starting set to $18$ and the maximum walltime for each iteration set to $T_{n,N}$, according to the number $n$ of vertexes of the graph instance and the targeted dimensionality $N$.
Figure \ref{fig:ffepochs} reports some statics on the first time the \textit{DEN} model find a feasible embedding for each trial, grouping the result by $n$ and separately for dimensionality $N$. You can notice that, due to the augmented dimensionality, it is easier for the \textit{DEN} model to find solutions in the $3D$ space. On average, the first feasible embedding is found earlier along the epochs. Moreover, as $n$ increases, it becomes more difficult to solve the CUDG problem. As we can see, for $n=70$, only one graph was successfully embedded within the $3000$ epochs in a $2D$ space.

\begin{figure}[ht!]
\centerline{\includegraphics[width=\linewidth]{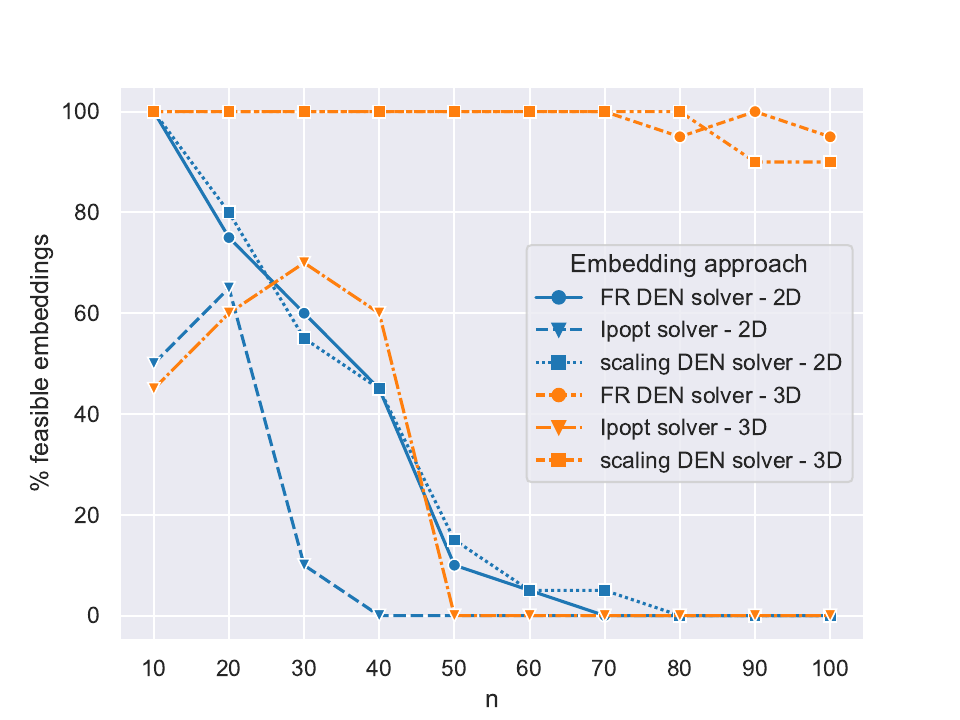}}
\caption{Percentage of feasible embeddings obtained, on the whole graphs dataset. The \textit{DEN} model outperforms the \textit{Ipopt} solver for both the $N=2$ and $N=3$ cases. There is no clear best choice concerning the initialization methods (\textit{Fruchterman-Reingold, i.e. FR}, or \textit{scaling}). As in the case of the first feasible embedding (Fig. \ref{fig:ffepochs}), the increment in dimensionality allows for finding more easily feasible embeddings, despite the increment in the models' parameters.}
\label{fig:perc_emb}
\end{figure}

This result is even better represented in Fig. \ref{fig:perc_emb} which shows the percentage of feasible embeddings retrieved with each optimization approach. In the case of the \textit{DEN} solver, the results achieved through the two initialization are distinguished. To better prove the advantage of the \textit{DEN} solver, we allowed the \textit{Ipopt} solver to exploit higher computational times. More precisely, we allowed for $10T_{n,N}$ (accordingly to $n$ and $N$, it could be more than half an hour). Nevertheless, it still did not provide feasible solutions for the instances that were not solved within $T_{n,N}$ walltime.

\begin{figure*}[ht!]
\centerline{\includegraphics[width=1.\linewidth]{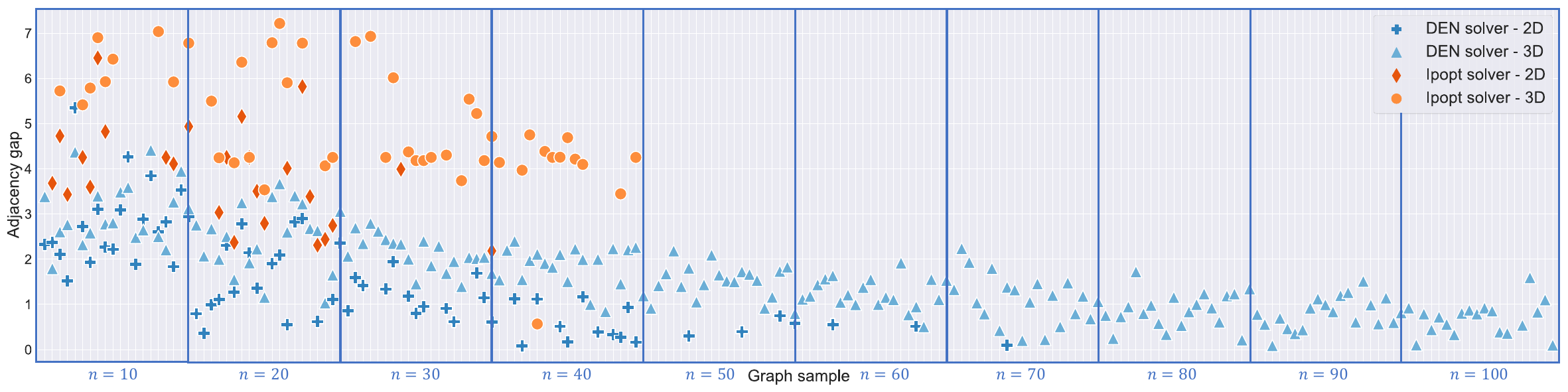}}
\caption{Comparison of the largest \textit{adjacency gap} values obtained with the different solvers, for all the graph samples in the dataset. The blue rectangles group the graphs' instances by $n$.}
\label{fig:gap}
\end{figure*}

Fig. \ref{fig:gap} illustrates instead the results concerning the maximization of the \textit{adjacency gap}. The \textit{DEN}-based optimization outperforms the classical optimization when the feasibility of the embedding is the goal of major importance. On the other hand, when the \textit{Ipopt}-based solver finds feasible embeddings, they correspond to a greater \textit{adjacency gap}. Possibly, more sophisticated optimization strategies on the $\alpha$ parameter in the \textit{ELF}, combined with an increment of $E$, could overcome this issue.

A final observation concerns the choice of the hyperparameters. Up to the experiment results, there is no specific combination of value for the learning rate $lr$ and the dropout probability $p_{drop}$ that provides a higher success rate for the CUDG solution. The retrieval of an embedding seems independent of those hyperparameters.

\section{Conclusion}\label{sec:conclusion}
This paper presents a novel neural-enhanced optimization framework that addresses a non-convex NP-hard optimization problem, \textit{i.e.}, the constrained unit disk graph problem. It arises from several real-world applications, such as QUBO problems' embedding for neutral-atoms-based quantum hardware.\\ The proposed \textit{distance encoder network} (\textit{DEN}) model combined with the \textit{embedding loss function} (\textit{ELF}) can find feasible embeddings for a larger set of graphs than the classical solver \textit{Ipopt}. Nonetheless, better embeddings could still be accomplished by improving the \textit{adjacency gap} optimization. Overcoming the limitation concerning \textit{adjacency gap} maximization will be the subject of further work, together with a deeper study on hyper-parameter settings and convergence analysis. Moreover, modified versions of the \textit{DEN} model and \textit{ELF} will be targeted to provide a more GPU-friendly implementation increasing the mini-batch size during the \textbf{training phase}.\\
A final observation concerns the generality of the approach. The \textit{DEN} model and the \textit{ELF} pursue the computation and optimization of Euclidean distances. Yet, custom modifications can supply outputs of interest for other optimization problems in a similar optimization framework. On this side, we can remark that the definition of the \textit{ELF} function is sufficiently general to model inequality constraints, and the activation functions combined with proper fixed-weight settings allow for representations of objective functions and constraints.


\bibliographystyle{IEEEtran}
\bibliography{icsa.bib}

\end{document}